\documentclass[aps,preprint,nofootinbib,showpacs,showkeys]{revtex4}
\usepackage{graphicx,amsmath}
\begin{document}
\title{Belle $J/\psi+\eta_c$ anomaly and a very light scalar boson}
\author{Kingman Cheung$^1$ and Wai-Yee Keung$^2$}
\affiliation{$^1$
Department of Physics and NCTS, National Tsing Hua University,
Hsinchu, Taiwan R.O.C. \\
$^2$ Department of Physics, University of Illinois at Chicago, 
Chicago IL 60607-7059, U.S.A.}
\date{\today}

\begin{abstract}
The large  rate of the $J/\psi+\eta_c$ production observed 
by the Belle collaboration has 
posed a serious challenge to our understanding of the quarkonium. 
We examine a scenario that there exists a  light scalar boson
around 5 GeV,  with an enhanced
coupling solely to the charm  quark.
Such a scenario would explain the Belle anomaly. 
It also predicts a significant increase in $J/\psi$ plus
open charm pair production.  
An immediate test for the scalar boson is to look for a peak around 5 GeV 
in the recoil mass spectrum of $J/\psi$, because the scalar boson can also be 
produced in association with the $J/\psi$.
Finally, we also point out that the process 
$e^+ e^- \to H \gamma$ is sizable for observation.
\end{abstract}
\maketitle

\section{Introduction}

The best understood QCD system is perhaps the heavy quarkonium, made up of
heavy quark and antiquark.  Known systems include charmonium,
bottomonium, and $B_c$ mesons.  We thought we understand these 
systems, but surprises often come up as more and more data become 
available.  A well-known incident was the surplus of $J/\psi$ produced
at high transverse momentum at the Tevatron \cite{cdf}.  
The color-singlet model \cite{singlet} missed
the cross section by more than an order of magnitude.  The color-octet
mechanism \cite{octet} and subsequently the nonrelativistic QCD model 
\cite{nrqcd} were formulated to explain the surplus.  

Recently, the Belle Collaboration observed double charmonium production,
 namely, $J/\psi+\eta_c$ at $\sqrt{s}=10.6$ GeV \cite{belle}.   
The production rate is anomalous \cite{belle,pakhlov,talk}
\begin{eqnarray}
\sigma(e^+ e^- \to J/\psi \; \eta_c ) &\times& B(\eta_c \to 2 \;{\rm charged})
\nonumber \\
&& \hspace{-0.3in} = 
( 0.033 \;^{+0.007}_{-0.006} \pm 0.009 ) \; {\rm pb} \,, \nonumber 
\end{eqnarray}
which is roughly an order of magnitude larger than any QCD
predictions \cite{kaoru}.  It poses a strong threat to our understanding of
quarkonium.  However, one has to note that the $J/\psi$ is detected
via its leptonic decay while $\eta_c$ is inferred from the recoil mass 
spectrum of the $J/\psi$.   Therefore, the $\eta_c$ is actually not identified,
and perhaps it could be a $J/\psi$, or something else.  
Based on this argument,
Bodwin, Lee, and Braaten \cite{bbl} argued that Belle in fact observed
double $J/\psi$'s, which are produced via double virtual photons.  However,
the further analysis by Belle does not seem to support this idea 
\cite{pakhlov}.
Thus, the anomalous production remains one of the  challenging  problems in
QCD, and arouses speculations. 
For example, a heavy glueball of a mass very near $m_{\eta_c}$ 
with a rather large matrix amplitude has been proposed\cite{Brodsky:2003hv} 
to account for the excess.

In this Letter, we study a possible scenario of  
the existence of a  light scalar boson $H$,
which has an enhanced coupling to the 
charm quarks.  It contributes to both the $J/\psi +\eta_c$ production, and
$J/\psi$ plus open charm pair production.  We show the allowed parameter
space region of the mass of this scalar and the coupling to charm, fitted
to the Belle data.  We find that a coupling constant $g_H\approx 1$ and 
$m_H\approx 5$ GeV can fit the data well.
An immediate test for this light scalar boson is to look for a peak around 
5 GeV in the recoil mass spectrum of $J/\psi$, because this scalar can also
be produced directly in association with a $J/\psi$.
So far, the Belle collaboration has only shown the recoil mass spectrum 
with very fine bin size up to 4 GeV \cite{belle,pakhlov,talk}, 
where $\eta_c, \chi_{c0}$, and $\eta_c(2S)$ can be seen.
However, the bin size in the region above 4 GeV is too large.
Therefore,
we urge the Belle collaboration to scan with very fine steps in the
region above 4 GeV.
(Even with a large bin size, we notice that there may be something peculiar 
at around 5.2--5.4 GeV bins \cite{belle,pakhlov}.)
Finally, we point out that another process
$e^+ e^- \to H \gamma$, which is negligible within the SM, becomes 
sizable in the our scenario.
This is another interesting test for the scenario.


\section{$e^+ e^- \to J/\psi \eta_c$}

Suppose the interaction of the scalar boson $H$ with the charm-quark 
pair is given by 
\begin{equation}
{\cal L} = - g_H \bar c \, c \, H \;,
\end{equation}
with a coupling constant $g_H$ that will be determined later.

Assuming that the scalar decays into a $c \bar c$ pair only, the 
width of $H$ is given by
\begin{equation}
\Gamma_{H} = \frac{3}{8\pi} g_H^2 m_H \left[ 1- \frac{4 m_c^2}{m_H^2} 
           \right ]^{3/2} \;.
\end{equation}
For $m_H\sim 5$ GeV and $g_H \sim 1$, the $\Gamma_H \sim 0.06 m_H$, 
which is narrow compared with its mass but much wider than the widths of 
$J/\psi$ and $\eta_c$.   In the following, we use the Breit-Wigner 
prescription for the scalar boson propagator.

\begin{figure}[t!]
\centering
\includegraphics[width=3.1in]{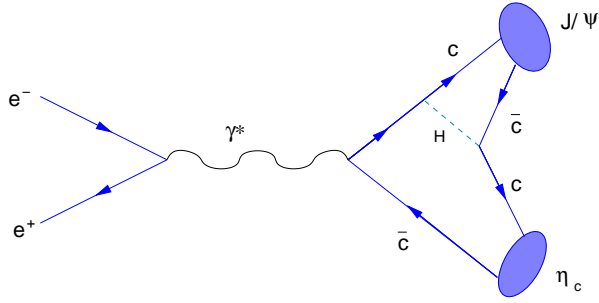}
\caption{\small One of the  four contributing Feynman diagrams for
$e^+ e^- \to J/\psi\, \eta_c$ via an intermediate scalar boson $H$.
\label{fey-1}}
\end{figure}

One of the contributing Feynman diagrams for 
$e^+ e^- \to J/\psi\, \eta_c$ is shown in Fig. \ref{fey-1}.
The sum of the amplitudes of the four contributing 
Feynman diagrams for $e^- (p_1) e^+ (p_2) \to J/\psi (k_1) \eta_c (k_2)$ 
is given by 
\begin{eqnarray}
i{\cal M} &=& - 4 g_H^2 e^2 Q_e Q_c \,\frac{ R_\psi(0) R_{\eta_c}(0)}{\pi}\,
\frac{1}{s^2} \, \frac{1}{s/4  - m_H^2 + i \Gamma_H m_H }  \nonumber \\
& \times &
\bar v(p_2) \gamma^\mu u(p_1)\,
\epsilon_{\mu\rho \lambda \sigma} k_2^\rho k_1^\sigma \epsilon^\lambda(k_1)
\end{eqnarray}
where $\epsilon(k_1)$ is the polarization 4-vector of the $J/\psi$, and
$s$ is the square of the center-of-mass energy of the collision. 
$R_\psi(0)$ and $R_{\eta_c}(0)$ are the wavefunctions at the origin,
with the normalization
$\int_0^\infty |R(r)|^2 r^2 dr=1$.
Here the center-of-mass energy $\sqrt{s}$ of the collision is at
$10.6$ GeV, and so we simply only take the intermediate photon
approximation (the intermediate $Z$ contribution is suppressed by
$s/m_Z^2 \sim 0.013$.)

The angular distribution of the process is given by
\begin{eqnarray}
\frac{d\sigma}{d \cos\theta} &=& \alpha^2 g_H^4 Q_e^2 Q_c^2 \,
\frac{ |R_\psi(0)|^2 |R_{\eta_c}(0)|^2}{\pi} \, \frac{1}{s^2} \nonumber \\
&& \hspace{-0.3in} \times
\frac{1}{ ( s/4  - m_H^2)^2 + \Gamma_H^2 m_H^2 } \, \beta^3 \,
 ( 1+  \cos^2\theta) \;,
\end{eqnarray}
where 
\[
\beta = \sqrt{ \left ( 1 - \frac{m_\psi^2}{s} - \frac{m_{\eta_c}^2}{s} 
\right)^2  - 4 \frac{m_\psi^2}{s} \frac{m_{\eta_c}^2}{s} } \;.
\]
Integrating the angle we obtain the total cross section
\begin{eqnarray}
\sigma &=& \frac{8}{3}
       \alpha^2 g_H^4 Q_e^2 Q_c^2 \,\frac{ |R_\psi(0)|^2 |R_\eta(0)|^2}
{\pi} \, \frac{1}{s^2} \nonumber \\
&& \hspace{-0.2in} \times
 \frac{1}{ ( s/4 - m_H^2 )^2 + \Gamma_H^2 m_H^2 } 
\,\beta^3
\label{cross}
\end{eqnarray}

\begin{figure}[t!]
\centering
\includegraphics[width=3.4in]{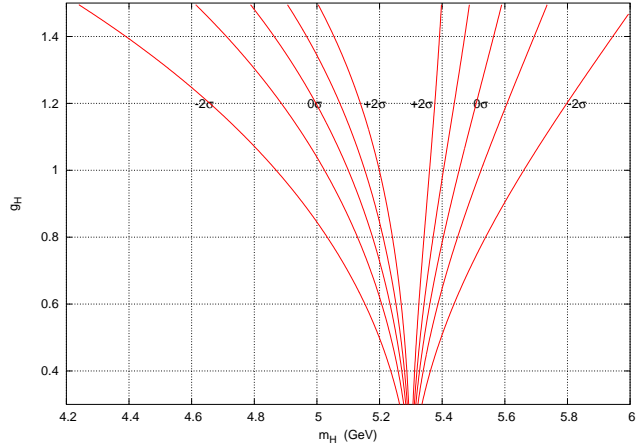}
\caption{\small 
The $\pm 1\sigma$ and $\pm 2\sigma$ bound for the $(m_H, g_H)$ plane due to 
the Belle data of $0.033 \pm 0.011$ pb.
\label{psieta}}
\end{figure}

Let us examine the allowed parameter space of $m_H$ and $g_H$ that can fit
the Belle cross section.  Combining the systematic and statistical errors
of the Belle data the $1\sigma$ range is $0.033 \pm 0.011$ pb.  We show the
contours of the central value, $\pm 1\sigma$, and $\pm2 \sigma$ in 
Fig. \ref{psieta}.  Taking $g_H=1$ the allowed central value ($2\sigma$ range)
for $m_H$ is
\begin{equation}
m_H = 5.10\,(4.87 - 5.20)\; {\rm GeV}\;\; {\rm or}\;\;
5.45\,(5.36 - 5.66)\; {\rm GeV} \;.
\end{equation}
Here we used $|R_\psi(0)|^2 = |R_\eta(0)|^2 =0.8\; {\rm GeV}^{3}$. 

It is straightforward to adapt our scenario to explain other peaks in the
recoil mass spectrum
between $3-3.8$ GeV, i.e, $\chi_{c0}$ and $\eta_c(2S)$ \cite{belle,pakhlov}.
  Since our
scenario only requires the $m_H$ at around $\sqrt{s}/2$, it would also 
explain the enhancement to the production of $J/\psi + (\eta_c(1S), 
\chi_{c0}, \eta_c(2S))$.  Note that our scenario would not give enhancement
to $J/\psi + (J/\psi, \chi_{c1}, \chi_{c2}, \psi(2S) )$ because
of the spin-parity-charge-conjugation of the photon propagator.  This is 
in agreement with the observation by Belle \cite{pakhlov}.

Note that if we choose a smaller $g_H$, the corresponding central value 
$m_H$ of the 
fit to the Belle data would become closer to $\sqrt{s}/2 = 5.3$ GeV, and the
$2\sigma$ range would also become smaller, as shown in Fig. \ref{psieta}. 
The smaller the value of $g_H$, the more fine-tuned value for $m_H$ that
the fit gives.  So in the following predictions of our scenario we simply
take $g_H=1$ and the results just scale as some powers of $g_H$.

If we replace the scalar $H$ by a pseudoscalar boson $A^0$, 
the diagram similar to Fig.~1 with $H$ replaced by $A^0$ 
gives vanishing amplitude in the static approximation. 
There is another diagram with the virtual $A^0$
bremsstrahlung off the charm quark 
turning into the $\eta_c$ state. However, such a contribution 
is important only when the mass $m_{A^0}$ is around $3-4$ GeV, which is
unfavorable based on  the constraint of the Belle search. 

The static approximation of using the wavefunction at the origin 
may not be so valid when the Higgs pole is involved, however our
simple calculation serves the purpose of estimating of the size of the
production rate due to the light Higgs boson $H$.

\section{$e^+ e^- \to J/\psi \, c \bar c$}

\begin{figure}[t!]
\centering
\includegraphics[width=3.4in]{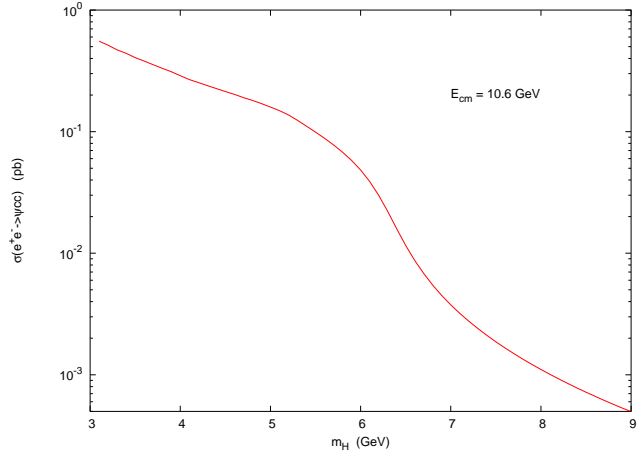}
\caption{\small 
The production cross section for $e^+ e^- \to J/\psi c \bar c$ Vs $m_H$.  The 
value of $g_H=1$ is chosen.
\label{psicc}}
\end{figure}

This process is very similar to the original $J/\psi + \eta_c$ production, 
except that the $c \bar c$ pair does not form a bound state but goes into
an open pair.  They will hadronize into $D$ mesons.  The calculation is 
straightforward, but however the result is not simple enough to be put 
here.  
We show the cross section verse $m_H$ at $\sqrt{s}=10.6$ GeV
in Fig. \ref{psicc}.  The cross section scales as $g_H^4$.
At $m_H\simeq 5$ GeV, $\sigma(e^+ e^- \to J/\psi c \bar c) \approx 0.15$ pb
for $g_H=1$.  Therefore, it contributes substantially to $J/\psi+$ open 
charm production.

\section{$e^+  e^-  \to \psi H$}

\begin{figure}[t!]
\centering
\includegraphics[width=3.4in]{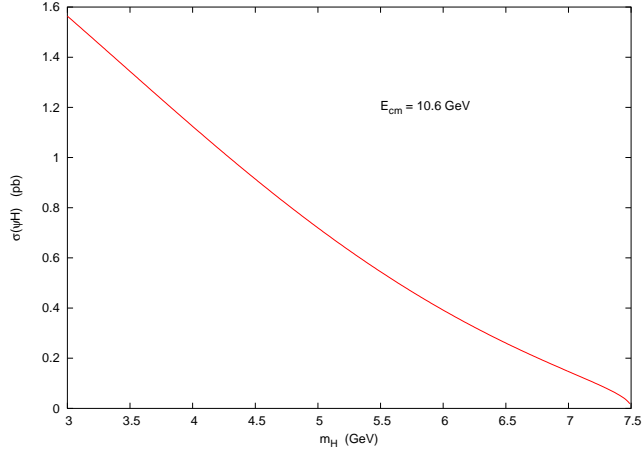}
\caption{\small 
The production cross section for $e^+ e^- \to J/\psi  H$ Vs $m_H$.  The 
value of $g_H=1$ is chosen.
\label{psih}}
\end{figure}

The sum of the two Feynman diagrams of the 
process $e^- (p_1) e^+ (p_2) \to \psi(k_1) H(k_2)$
is given by
\begin{eqnarray}
i{\cal  M} &=& 2 \sqrt{3} e^2 Q_e Q_c g_H \, 
  \frac{R_\psi(0)}{ \sqrt{\pi m_\psi} }\, \frac{1}{s} \, 
  \frac{1}{s - m_\psi^2 + m_H^2}  \nonumber \\
&& \hspace{-0.5in} \times  \bar v(p_2) \gamma_\mu u(p_1) \left [
    (s - m_H^2 + m_\psi^2) \epsilon^\mu(k_1) \,  - 2 k_2 \cdot \epsilon(k_1) 
      k_1^\mu \right ] \nonumber
\end{eqnarray}
%
The differential cross section is
\begin{eqnarray}
\frac{d\sigma}{d \cos\theta} &=& 6 \alpha^2 g_H^2 Q_e^2 Q_c^2 \,
  \frac{|R_\psi(0)|^2}{ m_\psi }\,  
    \frac{1}{(s - m_\psi^2 + m_H^2)^2} \,\beta \nonumber \\
&& \hspace{-0.2in} \times
  \,\left[
  4 \frac{m_\psi^2}{s}  +  \frac{1}{2} \beta^2 (1+ \cos^2 \theta ) \right ]
\end{eqnarray}
where 
\[
\beta = \sqrt{ \left ( 1 - \frac{m_\psi^2}{s} - \frac{m_H^2}{s} \right)^2 
 - 4 \frac{m_\psi^2}{s} \frac{m_H^2}{s} }
\]
The integrated cross section is 
\begin{equation}
\label{cross-a}
\sigma = 6 \alpha^2 g_H^2 Q_e^2 Q_c^2 
  \frac{|R_\psi(0)|^2}{ m_\psi } 
    \frac{1}{(s - m_\psi^2 + m_H^2)^2} \beta \left[
   8 \frac{m_\psi^2}{s} + \frac{4}{3} \beta^2 \right ]
\end{equation}
In Fig. \ref{psih}, we show the production cross section of $\sigma(\psi H)$
with $g_H=1$.  The cross section scales as $g_H^2$.
The scalar boson $H$ so produced is heavy enough to 
decay into an open  charm-quark pair.  Thus, it will give rise to open charm
pair production, with the pair around 5 GeV.  
If the Belle could scan the recoil mass spectrum of $J/\psi$ further up to
$5-6$ GeV, they should be able to see if this scalar boson $H$ exists.  This 
is the most immediate test for the scenario.

\section{$e^+ e^- \to H \gamma$}

\begin{figure}[t!]
\centering
\includegraphics[width=3.4in]{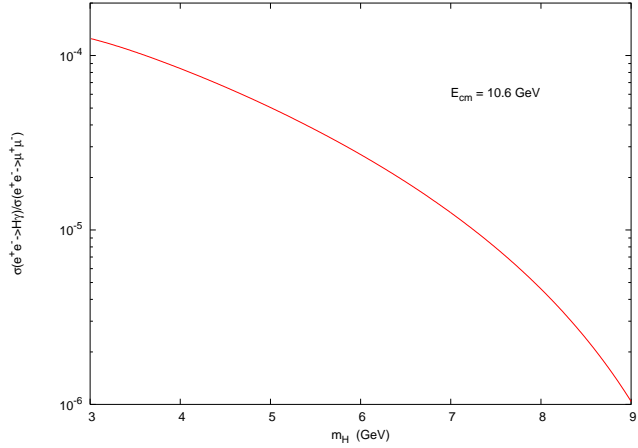}
\caption{\small 
The ratio of $\sigma( e^+ e^- \to H\gamma)/\sigma(e^+ e^- \to 
\mu^+ \mu^-)$ at $\sqrt{s}=10.6$ GeV for $m_H=3-9$ GeV.  The scenario is
$g_H=1$.
\label{hgamma}}
\end{figure}

In the SM, the leading contribution to the process 
$e^+ e^- \to H \gamma$ comes from the  one-loop diagrams 
with the $W$ boson and heavy fermions running in the loop.
In our scenario with $H$ coupling strongly to charm,
this channel could be sizable for detection.
The SM calculation was completed in Ref. \cite{leve}.  We use its formulas
with the charm-quark contribution only, and modify the SM coupling to our
enhanced coupling as follows:
\[
\sqrt{2} G_F m_c^2 \to g_H^2 \;.
\]
We show the ratio of $\sigma( e^+ e^- \to H\gamma)/\sigma(e^+ e^- \to 
\mu^+ \mu^-)$ at $\sqrt{s}=10.6$ GeV for $m_H=3-9$ GeV with $g_H=1$ 
in Fig. \ref{hgamma}.  The ratio scales as $g_H^2$.
Note that with the SM coupling strength $(g m_c/2m_W)$,
this ratio is very small, of order $O(10^{-9})$.  With the enhanced charm
coupling, the ratio is of order $5\times 10^{-5}$ for $m_H=5$ GeV.  Thus,
the cross section of $e^+ e^- \to H\gamma$ for $m_H=5$ GeV at $\sqrt{s}=
10.6$ GeV is about 40 fb, which will give thousands of 
events at Belle or BaBar with an integrated luminosity of 100 fb$^{-1}$.

\section{Discussion}

One may concern about the high energy analog of $ e^+ e^- \to 
c \bar c H$ production.  In high energy collisions, we would expect extra
$c \bar c c \bar c$ in the final state, with one of the $c\bar c$ pairs 
clustering at around 5 GeV.  To our knowledge there have not been 
any measurement on four-charm production, nor 
any searches explicitly looking for anomalous four-charm events at LEP or
at the Tevatron.  There were perhaps some mysterious
4-jet events recorded by ALEPH \cite{aleph}, 
but it had no explicit identification of charm in those events.  Also,
the search for four-charm events suffers severely from the weak charm 
identification, which is, to some extent, overlapped with the b-quark
identification.  Therefore, there is no explicit constraints on four-charm
production coming from high energy experiments.

Another possible source of constraints comes from Upsilon $\Upsilon(1S)$
 decays.  From the Particle Data Book \cite{PDG}, the branching ratio
$B(\Upsilon(1S) \to J/\psi + X) = 1.1 \pm 0.4 \%$.  
Given such a large error it is impossible to rule out our scenario,
because $J/\psi+\eta_c$, $J/\psi + H$, or $J/\psi +$ open charm
could only give a branching ratio below $0.4\%$.  
There is also 
a limit on $B(\Upsilon \to \gamma+X) < 3 \times 10^{-5}$, where $X$
is a pseudoscalar boson.  In our scenario, we have a scalar boson
and so the limit 
is not directly applicable. 
Even so, $B(\Upsilon \to H\gamma)/B(\Upsilon \to \mu^+ \mu^-)$ would 
be very similar to $\sigma(e^+ e^- \to H\gamma)/
\sigma(e^+ e^- \to \mu^+ \mu^-) \approx 5\times 10^{-5}$ 
(from Fig. \ref{hgamma}) without an explicit calculation.  It implies
$B(\Upsilon \to \gamma +H ) \approx 10^{-6}$, which is well below the
existing limit that we stated above.

The anomalous double charmonium $J/\psi + \eta_c$ production measured by
Belle posed a very strong challenge to our understanding of QCD.  We 
have investigated the scenario of a light scalar boson of mass
about $5$ GeV with a large coupling (of order the weak coupling or even larger)
to the charm quark.  We found that it
 could explain the $J/\psi + \eta_c$ anomaly, and
provided the much-needed $J/\psi$ plus open charm production.
It would also explain the enhancement to the production of 
$J/\psi + (\eta_c(1S), \chi_{c0}, \eta_c(2S) )$, but not 
to $J/\psi + (J/\psi, \chi_{c1}, \chi_{c2}, \psi(2S) )$ because
of the spin-parity-charge-conjugation of the photon propagator.  This is 
in agreement with the observation by Belle.
An immediate test for this light scalar boson is to scan 
the recoil mass spectrum of $J/\psi$ and look for a peak around 5 GeV, 
because this scalar can also
be produced directly in association with a $J/\psi$.
So far, the Belle collaboration has only shown the recoil mass spectrum 
with very fine bin size up to 4 GeV \cite{belle,pakhlov}, 
where $\eta_c, \chi_{c0}$, and $\eta_c(2S)$ can be seen.
However, the bin size in the region above 4 GeV is too large.
We, therefore, urge the Belle collaboration to
scan the recoil mass spectrum with very fine bin size above 4 GeV.
(Even with a large bin size, we notice that there may be something peculiar 
at around 5.2--5.4 GeV bins \cite{belle,pakhlov}, 
but this is only our wild guess.)
Finally, we have pointed out that another process
$e^+ e^- \to H \gamma$, which is negligible within the SM, becomes 
sizable in the new scenario of a light scalar boson, which couples strongly 
to charm.  This is another test for the scenario.

One may ask where this scalar boson comes from.  Such a light Higgs
may come from the general form of the multi-Higgs doublet sector. The
required enhanced coupling to the charm quark implies dedicated fine
tuning. It also requires additional fine tuning to kill the unwanted flavor 
changing neutral processes.
If we stay within the scope of the two-Higgs doublet sectors with the
quark couplings like those in  
the minimal supersymmetric standard model, 
the coupling strength of $H$ to charm
is $g m_c \cot\beta/2 m_W$.  Therefore, in order to have an effective coupling
$g_H \sim g$ we need a $\cot\beta \sim 2m_W/m_c \sim O(100)$.  Such a large
$\cot\beta$ is unnatural and perhaps causes a big problem to the top Yukawa
coupling.  We do not offer a good answer to the structure of the Higgs 
sector in this paper.

We thank Dr. Jungil Lee for discussion.

\end{document}